\documentclass[preprint,aps,prd,amsmath,byrevtex,prd,titlepage,
nofootinbib]{revtex4-1}

\usepackage{dcolumn}
\usepackage{amsmath}
\usepackage{amssymb}
\usepackage{bm}
\usepackage{hyperref}
\usepackage{graphicx}
\usepackage{slashed}

\newcommand{\beq}{\begin{equation}}
\newcommand{\eeq}{\end{equation}}
\newcommand{\eq}[1]{Eq.~(\ref{#1})}

\begin{document}

\title {Muon Loop Light-by-Light Contribution to Hyperfine Splitting in Muonium}
\author {Michael I. Eides}
%\altaffiliation[Also at ]{the Petersburg Nuclear Physics Institute,
%Gatchina, St.Petersburg 188300, Russia}
\email[Email address: ]{eides@pa.uky.edu, eides@thd.pnpi.spb.ru}
\affiliation{Department of Physics and Astronomy,
University of Kentucky, Lexington, KY 40506, USA\\
and Petersburg Nuclear Physics Institute,
Gatchina, St.Petersburg 188300, Russia}
\author{Valery A. Shelyuto}
\email[Email address: ]{shelyuto@vniim.ru}
\affiliation{D. I.  Mendeleyev Institute for Metrology,
St.Petersburg 190005, Russia}
%\date{}

\begin{abstract}
Three-loop corrections to hyperfine splitting in muonium generated by the gauge invariant sets of diagrams with muon and tauon loop light-by-light scattering blocks are calculated. These results complete calculations of all light-by-light scattering contributions to hyperfine splitting in muonium.

\end{abstract}

%\pacs{12.20Ds,31.30.jf,32.10.Fn,36.10.Ee}
%\keywords{hyperfine splitting}

\preprint{UK/14-01}

\maketitle

%\section{Introduction}

Calculation of the light-by-light (LBL) scattering contributions to hyperfine splitting in muonium has a long history. The nonrecoil contribution generated by the electron LBL scattering block was obtained in \cite{eks3,kn1,kn10}. Respective recoil contributions are enhanced by the large logarithm of the muon-electron mass ratio. Large logarithm squared contributions were calculated in \cite{eks89}, and single-logarithmic and nonlogarithmic terms were obtained only recently \cite{es2013_1,es2013_2}. The LBL scattering contributions due to other particles besides the electron also should be taken into account. The hadron LBL scattering contribution was calculated in \cite{ksv2008}. Below we present the results for the only remaining still uncalculated LBL scattering contributions to hyperfine splitting in muonium due to the virtual muon and tauon loops.

\begin{figure}[htb]
\includegraphics
[height=3cm]
{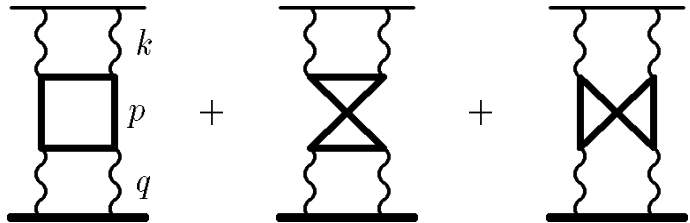}
\caption{\label{lblrec}
Diagrams with the muon (tauon) light-by-light scattering block}
\end{figure}

The general expression for the muon loop LBL  scattering contribution to HFS  in Fig. \ref{lblrec} is similar to the respective electron loop contribution (see, e.g., \cite{es2013_1,es2013_2}), and can be written in the form

\beq
\Delta E=\frac{\alpha^2(Z\alpha)}{\pi^3}\frac{m}{M}E_FJ,
\eeq

\noindent
where $m$ is the electron mass, $M$ is the muon mass, $Z=1$ is the muon charge in terms of the positron charge used for classification of different contributions, the Fermi energy is defined as ($m_r$ is the reduced mass)

\beq
E_F=\frac{8}{3}(Z\alpha)^4\frac{m}{M}\left(\frac{m_r}{m}\right)^3mc^2,
\eeq

\noindent
and $J$ is a dimensionless integral

\beq \label{jint}
J=-\frac{3M^2}{128}\int \frac{d^4k}{i\pi^2 k^4}
\left(\frac{1}{k^2+2mk_0}+\frac{1}{k^2-2mk_0}\right)T(k^2,k_0).
\eeq

\noindent
The dimensionless function $T(k^2,k_0)$ is a sum of the ladder and crossed diagrams contributions in Fig. \ref{lblrec}

\beq
T(k^2,k_0)=2T_L(k^2,k_0)+T_C(k^2,k_0).
\eeq

\noindent
Explicit expressions for the functions $T_L(k^2,k_0)$ and $T_C(k^2,k_0)$ can be obtained  by the substitution $m\to M$, $q_\mu\to k_\mu$  from the respective formulae in \cite{es2013_2}, where these functions were calculated in the case of the electron LBL scattering block.

Only the even in $k_0$ terms in the function $T(k^2,k_0)$ contribute to the integral in \eq{jint}. After rescaling of the integration momentum $k\to kM$, the Wick rotation, and  symmetrization of the function $T(k^2,k_0)$ with respect to $k_0$, $T(k^2,k_0)\to T(k^2,k_0^2)$, the integral in \eq{jint} turns into

\beq \label{wickene}
J=\frac{3}{32\pi}\int_0^\infty \frac{dk^2}{k^2}\int_0^\pi d\theta\sin^2\theta\frac{T(k^2,\cos^2\theta)}{k^2+16\mu^2\cos^2\theta},
\eeq

\noindent
where we have parameterized the Euclidean four-vectors as $k_0=k\cos\theta$, $|\bm k|=k\sin\theta$, $\mu=m/(2M)$, and the function  $T(k^2,\cos^2\theta)$ is the same function as in \eq{jint} but symmetrized with respect to $k_0$ and with the Wick rotated momenta. The dimensionless function $T(k^2,\cos^2\theta)$ after rescaling depends on the dimensionless momentum $k$ and does not contain any parameters with dimension of mass.

We are looking for the $\mu$-independent contributions generated by the integral in \eq{wickene}. The term with $\mu^2$ in the denominator is irrelevant at large $k$, and the integral is convergent at large $k$ due to ultraviolet convergence of all diagrams with the LBL insertions.  The case of small integration momenta is more involved. Due to gauge invariance, the LBL block is strongly suppressed at $k\to0$, and we expect that the integral in \eq{wickene} remains finite even at $\mu=0$ zero. As a result of this finiteness the diagrams in Fig. \ref{lblrec} do not generate nonrecoil contributions to HFS in accordance with our physical expectations. However, small integration momenta convergence of contributions of individual diagrams at $\mu=0$ cannot be taken for granted, and we have to to consider separate entries in more detail. Using the explicit integral representations for the functions $T_{L}(k^2,\cos^2\theta)$ and $T_{C}(k^2,\cos^2\theta)$ (see \cite{es2013_2}) we find that these functions, and separate terms in the respective integral representations, decrease not slower than $k^2$ at $k^2\to0$ . The integral in \eq{wickene} is logarithmically divergent at $\mu=0$ if $T(k^2)\sim k^2$ when $k^2\to0$. This means that we cannot omit $\mu$ in \eq{wickene} calculating the integrals with those terms in $T(k^2)$ that decrease as $k^2$ when $k^2\to0$. To facilitate further calculations we represented the functions $T_{L}(k^2,\cos^2\theta)$ and $T_{C}(k^2,\cos^2\theta)$ in the form

\beq
T_{L}(k^2,\cos^2\theta)=T^{reg}_{L}(k^2,\cos^2\theta)
+T^{sing}_{L}(k^2,\cos^2\theta),
\eeq
\beq
T_{C}(k^2,\cos^2\theta)=T^{reg}_{C}(k^2,\cos^2\theta)
+T^{sing}_{C}(k^2,\cos^2\theta),
\eeq

\noindent
where the functions $T^{reg}$'s decrease faster than $k^2$  at small $k^2$, and the functions $T^{sing}$'s decrease as $k^2$  at small $k^2$.

In these terms the integral in \eq{wickene}  has the form

\beq
J=J^{reg}+J^{sing},
\eeq

\noindent
where

\beq \label{singregint}
J^{reg(sing)}=\frac{3}{32\pi}\int_0^\infty \frac{dk^2}{k^2}\int_0^\pi d\theta\sin^2\theta\frac{T^{reg(sing)}(k^2,\cos^2\theta)}{k^2+16\mu^2\cos^2\theta},
\eeq

\noindent
and

\beq
T^{reg(sing)}(k^2,\cos^2\theta)=2T^{reg(sing)}_{L}(k^2,\cos^2\theta)
+T^{reg(sing)}_{C}(k^2,\cos^2\theta).
\eeq

\noindent
We can safely let $\mu=0$ in the integral $J^{reg}$, what makes calculation of this integral straightforward. As a result we obtain

\beq \label{jreg}
J^{reg}=-2.146~35(5).
\eeq

Calculation of the integral $J^{sing}$ is more involved. The functions $T^{sing}_{L}(k^2,\cos^2\theta)$ and $T^{sing}_{C}(k^2,\cos^2\theta)$ decrease as $k^2$ at low $k^2$. As a result they generate logarithmic contributions to the momentum integral in \eq{wickene} that are cutoff at small $k\sim\mu$. We calculated the coefficients before the terms proportional to $k^2$  analytically and checked that these terms cancel in the sum  $2T^{sing}_{L}(k^2,\cos^2\theta)+T^{sing}_{C}(k^2,\cos^2\theta)$. This cancelation can be used to get rid of the parameter $\mu$ in the integral $J^{sing}$ in \eq{singregint}. To this end we write the momentum integral $J^{sing}$ as a sum of two integrals

\beq
J^{sing}=J^{sing<}+J^{sing>},
\eeq

\noindent
where integration over $k^2$ goes from zero to $1$ in the integral $J^{sing<}$, and it goes from $1$ to infinity in the integral $J^{sing>}$ . The separation point $k^2=1$ is arbitrary, the result for the integral $J^{sing}$ does not depend on its choice. We can safely let $\mu=0$ in  the integral $J^{sing>}$. To facilitate calculation of the integral $J^{sing<}$ we subtract from the integrand all terms proportional to $k^2$ at small $k$. Due to cancelation mentioned above this subtraction does not change the value of the integral. After the subtraction we can let $\mu=0$ before calculation of this integral as well. Calculating the integrals we obtain

\beq
J^{sing<}=0.174~47(2),          \qquad
J^{sing>}=1.129~51(3).
\eeq

\noindent
We have checked by direct calculations that the sum

\beq \label{jsing}
J^{sing<}+J^{sing>}=1.303~98(4).
\eeq

\noindent
does not depend on the arbitrary separation point.

Collecting the results in \eq{jreg} and \eq{jsing} we obtain

\beq
J=-0.842~4(1),
\eeq

\noindent
and finally

\beq \label{muonloopcon}
\Delta E=-0.842~4(1)\frac{\alpha^2(Z\alpha)}{\pi^3}\frac{m}{M}E_F
\approx-0.2274~\mbox{Hz}.
\eeq

Using the same methods as above we also calculated a tiny contribution to hyperfine splitting generated by the tauon LBL scattering block in Fig. \ref{lblrec}

\beq \label{tauonloopcon}
\Delta E_\tau=-0.003~58(1)\frac{\alpha^2(Z\alpha)}{\pi^3}\frac{m}{M}E_F
\approx-0.0010~\mbox{Hz}.
\eeq

\noindent
Combining the results in \eq{muonloopcon} and \eq{tauonloopcon} with the other LBL scattering contributions calculated earlier in \cite{eks3,kn1,kn10,eks89,es2013_1,es2013_2,ksv2008} we obtain the total contribution of the LBL scattering block to hyperfine splitting in muonium

\beq
\begin{split}
\Delta E=
&\frac{\alpha^2(Z\alpha)}{\pi}(1+a_\mu)E_F[-0.472~514~(1)]
\\
&+\frac{\alpha^2(Z\alpha)}{\pi^3}\frac{m}{M}E_F
\left[\frac{9}{4}\ln^2{\frac{{M}}{m}}
+ \biggl(-3\zeta{(3)}- \frac{2\pi^2}{3} + \frac{91}{8}\biggr) \ln{\frac{{M}}{m}} +5.152~5(1)\right]
\\
&+\Delta E_{hadr}+\Delta E_\tau
\approx-240.2~\mbox{Hz},
\end{split}
\eeq

\noindent
where $\Delta E_{hadr}=-0.0065$ Hz is the hadronic contribution \cite{ksv2008}.

Completion of calculations of all LBL scattering contributions to hyperfine splitting in muonium brings us one step closer to the final goal of reducing the theoretical error of the hyperfine splitting in muonium below 10 Hz \cite{egs2007}. Hopefully the new result in \eq{muonloopcon} will find applications in the new high accuracy measurement of the muonium hyperfine splitting planned at J-PARC, Japan \cite{shimomura}.

%\begin{acknowledgments}

This work was supported by the NSF grant PHY-1066054. The work of V. S. was also supported in part by the RFBR grant 12-02-00313 and by the DFG grant GZ: HA 1457/7-2.

%\end{acknowledgments}

\end{document}